\documentclass[aps,prl,preprint,groupedaddress]{revtex4-1}
\usepackage{graphicx}
\usepackage{amsmath}
\usepackage{amssymb}
\usepackage{graphicx}
\usepackage{mathrsfs}
\usepackage{lineno}
\usepackage{bbm}

\begin{document}

\title{Path integral approach to generating functions for multistep post-transcription and post-translation processes and arbitrary initial conditions.}

\author{Jaroslav Albert}
\affiliation{Universit\'e Catholique de Louvain, Institute of Information and Communication Technologies, Electronics and Applied Mathematics, Louvain-la-Neuve, Belgium}
\email{jaroslavalbert81@gmail.com}

\begin{abstract}
Stochastic fluctuations in the copy number of gene products have perceivable effects on the functioning of gene regulatory networks (GRN).
The Master equation (ME)
provides a theoretical basis for studying such effects. However, solving the ME can be a task that ranges from
simple to difficult to impossible using conventional methods. Therefore, discovering new techniques for solving the ME is
an important part of research on stochastic GRN. In this paper, we present a novel approach to obtaining the generating function (GF),
which contains the same information as the ME, for
a one gene system that includes multi-step post-transcription and post-translation processes. The novelty of the approach lies in
the separation of the mRNAs from proteins. The GF for the mRNAs is obtained using a formalism involving operators and vector states. 
Using the same formalism, the
GF for the proteins is solved for a particular path taken by all mRNAs in the time-copy number plane;
then, the GF is summed over all possible paths. We prove a theorem that shows the summation of all paths
to be equivalent to an equation similar to the ME for the mRNAs. 
On a system with six gene products in total and randomly selected initial conditions, we confirm the validity of our results by
comparing them with Gillespie simulations.
\end{abstract}

\maketitle

\section{Introduction}

Stochastic fluctuations give rise to cell-to-cell differences in copy numbers of gene products, such as mRNA and protein.
Sometimes these differences are insignificant; other times they lead
to major shifts in phenotype \cite{Blake}.
Therefore, understanding the impact of stochastic fluctuations is an important endeavor in the field of systems biology.

There exist numerous methods for modeling stochastic gene expression. Some of them are entirely numerical, such as
the Gillespie algorithm \cite{Gillespie} and its derivatives \cite{Gibson}, \cite{Gillespie2}, \cite{Cao}, \cite{Cao2}, \cite{Cao3}; 
others are hybrids of the Gillespie algorithm and the Master equation
\cite{Burrage}, \cite{Jahnke}, \cite{Albert}, \cite{Albert12}, \cite{Duso}, \cite{Alfonsi}, \cite{Kurasov}; while the rest facilitate either exact or 
approximate analytic solutions to the Master equation \cite{Jahnke2}, \cite{Albert2}.
Analytic solutions are of great value because they provide a more direct insight into the system's behavior, and/or allow for
fast exploration of the system's parameter space.
However, solving the Master equation analytically has proven possible only for systems that are too simple and hence not very interesting. For more complex systems,
the strategy is usually to find techniques that lead to approximate (but analytic) solutions of either the Master equation or
the generating function \cite{Shahrezaei} and \cite{Pendar}.

In this paper we present a novel approach to obtaining the generating function of a one-gene system comprising
of partially and fully processed mRNA and protein. We make the observation that the Master equation can be reduced to include
the mRNAs only, and that a similar reduction can be achieved also for the proteins but only for a 
specific path in the time-copy number space of the fully processed mRNAs. We show, by proving a theorem involving
a sum of all paths of the partially and fully processed mRNA, that obtaining the generating 
function for arbitrary initial conditions can be reduced to solving a set of $M$ ordinary differential equations,
where $M$ is the number of post-transcription processes. Solving these equations numerically, and with the help of Cauchy's contour theorem, we compute the
probability distributions at different times for the fully processed protein. Also, we compute the first four moments
for the fully processed protein starting with a randomly generated set of initial values for all the gene products.
We demonstrate the validity of our approach by comparing our results with Gillespie simulations.
We conclude by proposing a further use of the above-mentioned theorem in, e. g. reducing dimensionality in a multi-gene system.

\section{The Master equation}

The system we consider comprises of these reactions:
\begin{eqnarray}\label{reactions}
\phi&\xrightarrow{\makebox[1cm]{$r$}}&m_1\nonumber\\
m_M&\xrightarrow{\makebox[1cm]{$d$}}&\phi\nonumber\\
m_i&\xrightarrow{\makebox[1cm]{$a_i$}}&m_{i+1}\,\,\,\,\,\,\,\,i=1,...,M-1\nonumber\\
m_i&\xrightarrow{\makebox[1cm]{${\bar a}_i$}}&m_{i-1}\,\,\,\,\,\,\,\,i=2,...,M\nonumber\\
m_M&\xrightarrow{\makebox[1cm]{$K$}}&m_M+n_1\nonumber\\
n_N&\xrightarrow{\makebox[1cm]{$q$}}&\phi\nonumber\\
n_i&\xrightarrow{\makebox[1cm]{$b_i$}}&n_{i+1}\,\,\,\,\,\,\,\,i=1,...,N-1\nonumber\\
n_i&\xrightarrow{\makebox[1cm]{${\bar b}_i$}}&n_{i-1}\,\,\,\,\,\,\,\,i=2,...,N,\nonumber\\
\end{eqnarray}
where $m_1$ is the copy number of freshly transcribed mRNAs, $m_i$, for $i=2,3,...,M$, are the copy numbers of mRNAs that
have undergone the first $(i=2)$, second $(i=3)$, etc. post-transcription process with $m_M$ being the copy number of fully processed
mRNA from which proteins can be translated. The same notation applies to the proteins: $n_1$ is the copy number of freshly transcribed proteins, and $n_N$
is the copy number of fully processed proteins.
For the reactions that change the variables $m_i$, $r$ is the transcription rate, $d$ is the degradation rate of a fully processed mRNA, and $a_i$ and ${\bar a}_i$
are the forward and backward reaction rates of the post-transcription processes, respectively. For the remaining reactions,
$K$ is the translation rate, $q$ is the degradation rate of a fully processed protein, and $a_i$ and ${\bar a}_i$
are the forward and backward reaction rates of the post-translation processes, respectively.
The master equation for this system reads
\begin{eqnarray}\label{CME_total}
\frac{\partial}{\partial t}P({\bf m},{\bf n},t)&=&
r(t)[P(m_1-1)-P]+d[(m_M+1)P(m_M+1)-m_MP]\nonumber\\
&&+\sum_{i=1}^{M-1}a_i[(m_i+1)P(m_i+1,m_{i+1}-1)-m_iP]\nonumber\\
&&+\sum_{i=1}^{M-1}{\bar a}_{i+1}[(m_{i+1}+1)P(m_i-1,m_{i+1}+1)-m_{i+1}P]\nonumber\\
&&+Km_M[P(n_1-1)-P]+q[(n_N+1)P(n_N+1)-n_NP]\nonumber\\
&&+\sum_{i=1}^{N-1}b_i[(n_i+1)P(n_i+1,n_{i+1}-1)-n_iP]\nonumber\\
&&+\sum_{i=1}^{N-1}{\bar b}_{i+1}[(n_{i+1}+1)P(n_i-1,n_{i+1}+1)-n_{i+1}P],
\end{eqnarray}
where $P({\bf m},{\bf n},t)$ is the joint probability of observing the sets of copy numbers ${\bf m}=(m_1,...,m_M)$
and ${\bf n}=(n_1,...,n_N)$.
For brevity,
we only write the argument(s) of $P$ on the right hand side of Eq. (\ref{CME_total}) explicitly if there is a change to
the said argument(s); e. g. instead of writing $P({\bf m},n_1,...,n_k+1,...,n_N,t)$, we write
$P(n_k+1)$. If there is no change to any argument, we merely write $P$.
In principal, Eq. (\ref{CME_total}) could be solved
numerically; however, even for moderate average mRNA and protein copy numbers, the dimension of the problem
might be too large for such a direct approach. For example, for $M=N=3$, $m_1=m_2=m_3\sim10$ and $n_1=n_2=n_3\sim100$,
the number of equations that need to be solved are of order $3^2\times10^3\times100^3=9\times10^9$.

\section{The generating function}
\subsection{Direct approach}

An alternative approach to the above problem is to obtain a generating function (GF), which contains as much information
about the system as the Master equation. If we let ${\bf s}$ be the set of all variables $(m_1,...,m_M,n_1,...,n_N)$,
then the GF is defined as
\begin{equation}\label{GF}
F({\bf y},t)=\sum_{{\bf s}}\prod_{k=1}^{M+N}y_k^{s_k}P({\bf s},t).
\end{equation}
The probability distribution, $P(s_k,t)$, for the variable $s_k$ is related to $F({\bf s},t)$ through
this expression:
\begin{equation}\label{PF}
P(s_k,t)=\frac{1}{s_k!}\left[\frac{\partial^{s_k} F({\bf y},t)}{\partial y^{s_k}}\right]_{{\bf y}=0}.
\end{equation}
The GF can also be used to obtain statistical moments:
\begin{equation}\label{moments}
\langle s_k^l\rangle=\left[\left(y_k\frac{\partial}{\partial y_k}\right)^lF({\bf y},t)\right]_{y_1=1,...,y_{M+N}=1}=\sum_{{\bf s}}s_k^lP({\bf s},t).
\end{equation}
To obtain an equations from the GF, we must multiply Eq. (\ref{CME_total}) by the product $\prod_{i=1}^{M+N}y_i^{s_i}$ and sum
over ${\bf s}$. The result is a partial differential equation of the form \cite{Walczak}
\begin{eqnarray}\label{GF_Eq}
\frac{\partial F}{\partial t}&=&
r(t)(y_1-1)F-d(y_M-1)\frac{\partial F}{\partial y_M}\nonumber\\
&+&\sum_{i=1}^{M-1}a_i(y_{i+1}-y_i)\frac{\partial F}{\partial y_i}
+\sum_{i=1}^{M-1}{\bar a}_{i+1}(y_i-y_{i+1})\frac{\partial F}{\partial y_{i+1}}\nonumber\\
&+&K(y_{M+1}-1)y_M\frac{\partial F}{\partial y_M}-q(y_{M+N}-1)\frac{\partial F}{\partial y_{M+N}}\nonumber\\
&+&\sum_{i=1}^{M-1}b_i(y_{M+i+1}-y_{M+i})\frac{\partial F}{\partial y_{M+i}}
+\sum_{i=1}^{M-1}{\bar b}_{i+1}(y_{M+i}-y_{M+i+1})\frac{\partial F}{\partial y_{M+i+1}}\nonumber\\.
\end{eqnarray}
For the system at hand, even this equation is very difficult to solve, especially for arbitrary initial conditions.
In the next section, we show an alternative approach for obtaining the GF.

\subsection{Path integral approach}

We begin by noticing that the stochastic evolution of the set ${\bf m}=(m_1,...,m_M)$ is independent of the set ${\bf n}=(n_1,...,n_N)$.
To see this, we sum both sides of Eq. (\ref{CME_total}) over ${\bf n}$ to obtain
\begin{eqnarray}\label{CME_mRNA}
\frac{\partial}{\partial t}P({\bf m},t)&=&
r(t)[P(m_1-1)-P]+d[(m_M+1)P(m_M+1)-m_MP]\nonumber\\
&&+\sum_{i=1}^{M-1}a_i[(m_i+1)P(m_i+1,m_{i+1}-1)-m_iP]\nonumber\\
&&+\sum_{i=1}^{M-1}{\bar a}_{i+1}[(m_{i+1}+1)P(m_i-1,m_{i+1}+1)-m_{i+1}P].\nonumber\\
\end{eqnarray}
If the initial probability, $P({\bf m},0)$, is Poisson, then
$P({\bf m},t)$ is also Poisson:
\begin{equation}\label{PoissonM}
P({\bf m},t)=\prod_{i=1}^M\frac{\gamma_i^{m_i}}{m_i!}e^{-\gamma_i},
\end{equation}
where the vector ${\pmb\gamma}$ satisfies
\begin{equation}\label{gamma}
\frac{d{\pmb\gamma}}{dt}={\bf S}{\pmb\gamma}+{\bf r},
\end{equation}
in which
\begin{eqnarray}
{\bf S}=
\left[\begin{array}{cccccccccc}
-a_1 & {\bar a}_2 &  &  &  &  & &  \\
a_1 & -(a_2+{\bar a}_2) & {\bar a_3} &  &  &  &  & & &\\
&   & {\huge\text{.}} &   &   &   &   &   &   &\\
&   &                         &   &    & {\huge\text{.}} &   &   &   &\\
&   &                         &   &    &                         &    &  {\huge\text{.}}  &   &\\
&   &   &   &   &   &   &   &   &\\
&   &   &   &   &   &   &   &   &\\
&   &   &   &   &   &   & a_{M-2}  & -(a_{M-1}+{\bar a}_{M-1}) &  {\bar a}_M\\
&   &   &   &   &   &   &  & a_{M-1} &  -(a_M+d)\nonumber
\end{array}\right],
\end{eqnarray}
and
\begin{eqnarray}
{\bf r}=
\left[\begin{array}{c}
r(t)\\
0\\
{\huge\text{.}}\\
{\huge\text{.}}\\
{\huge\text{.}}\\
0\nonumber
\end{array}\right],
\end{eqnarray}
The solution to Eq. (\ref{gamma}) is
\begin{equation}\label{gamma_sol}
\gamma_k=\int_0^tf_k(t-t')r(t')dt'+\sum_{i=1}^M\sum_{j=1}^MU_{ki}U_{ij}^{-1}e^{S_it}\gamma_j(0),
\end{equation}
where $S_i$ is the $i^{th}$ eigenvalue of ${\bf S}$, $U_{ij}$ is the unitary matrix that diagonalizes
${\bf S}$, i. e. ${\bf U}^{-1}{\bf S}{\bf U}=\delta_{ij}S_j$, and
\begin{equation}\label{f_def}
f_k(t-t')=\sum_{i=1}^MU_{ki}U_{i1}^{-1}e^{S_i(t-t')}.
\end{equation}

Before we attempt to generalize the solution in Eq. (\ref{PoissonM}) for arbitrary initial conditions, it will prove useful to reformulate
the ME (\ref{CME_mRNA})
in terms of a tensor product of basis vectors $|{\bf m}\rangle=|m_1\rangle|m_2\rangle...|m_M\rangle$ and
its transpose $\langle {\bf m}|=\langle m_1|\langle m_2|...\langle m_M|$, and operators $A_i$, $A_i^{+}$ and $A_i^{+}A_i$, whose action on  $|{\bf m}\rangle$
is as follows (see review \cite{Walczak} and references therein):
\begin{eqnarray}\label{AAplus}
A_i|{\bf m}\rangle&=&m_i|m_1\rangle|m_2\rangle...|m_i-1\rangle...|m_M\rangle,\nonumber\\
A_i^{+}|{\bf m}\rangle&=&|m_1\rangle|m_2\rangle...|m_i+1\rangle...|m_M\rangle,\nonumber\\
A_i^{+}A_i|{\bf m}\rangle&=&m_i|{\bf m}\rangle.
\end{eqnarray}
In this notation, the ME (\ref{CME_mRNA})
can be written as
\begin{equation}\label{psi}
\frac{d}{dt}|\psi\rangle=H|\psi\rangle,
\end{equation}
where
\begin{equation}
|\psi\rangle=\sum_{\bf m}P({\bf m},t)|{\bf m}\rangle.
\end{equation}
and
\begin{eqnarray}
H&=&r(t)[A_1^+-1]+d[A_M-A_M^+A_M]\nonumber\\
&&+\sum_{i=1}^{M-1}a_i[A_iA^+_{i+1}-A_i^+A_i]\nonumber\\
&&+\sum_{i=1}^{M-1}{\bar a}_{i+1}[A_{i+1}A_i^+-A_{i+1}^+A_{i+1}].
\end{eqnarray}
The formal solution to Eq. (\ref{psi}) can be written as
\begin{equation}\label{psisol}
|\psi\rangle=e^{H(t_{L-1})dt}e^{H(t_{L-2})dt}\,.\,.\,.\,e^{H(t_0)dt}|\psi(0)\rangle,
\end{equation}
where 
\begin{equation}\label{InitialState}
|\psi(0)\rangle=\sum_{\bf m}P({\bf m},0)|{\bf m}\rangle
\end{equation}
is the initial state, $t_0=0$, $t_1=dt$, etc. and $Ldt=t$. 
Using the orthogonality
relations $\langle m_i|m'_j\rangle=\delta_{m_im'_i}\delta_{ij}$, we obtain the probability of observing ${\bf m}$ at time $t$
by multiplying Eq. (\ref{psisol}) by $\langle {\bf m}|$:
\begin{equation}\label{psisol2}
P({\bf m},t)=\langle {\bf m}|\psi\rangle=\langle {\bf m}|e^{H(t_{L-1})dt}e^{H(t_{L-2})dt}\,.\,.\,.\,e^{H(t_0)dt}|\psi(0)\rangle,
\end{equation}
or, in a short hand notation
\begin{equation}\label{psisol2}
P({\bf m},t)=\langle {\bf m}|\psi\rangle=\langle {\bf m}|G(t)|\psi(0)\rangle,
\end{equation}
where
\begin{equation}\label{psisol2}
G(t)=Te^{\int_0^tH(t')dt'}
\end{equation}
and $T$ is the time ordering operator. The operator $G(t)$ can also act on
a complex Poisson state defined by 
\begin{equation}
|{\bf z}\rangle=\sum_{\bf m}\prod_{i=1}^M\frac{z_i^{m_i}}{m_i!}e^{-z_i}|{\bf m}\rangle,
\end{equation}
where ${\bf z}=(z_1,...,z_M)$ is a set of complex numbers. The operator $G(t)$ acts on the state $|{\bf z}\rangle$ by
evolving the variables ${\bf z}$, i. e. ${\bf z}\rightarrow {\bf z}(t)$, according to Eq. (\ref{gamma}).
The solution of Eq. (\ref{gamma}), ${\bf z}(t)$, is given by Eq. (\ref{gamma_sol}) but
with $\gamma_j(0)$ replaced by $z_j=z_j(0)$.

To solve the Master equation (\ref{CME_mRNA}) for arbitrary initial conditions,
we need to know how $G(t)$ transforms the state $|{\bf m}\rangle$. 
Since we know how it acts on the state $|{\bf z}\rangle$, we must express $|{\bf m}\rangle$ in terms of $|{\bf z}\rangle$.
This can be done for each state $|m_j\rangle$ via the identity
\begin{equation}\label{zIntegral}
|m\rangle=\oint dz\frac{m!}{2\pi iz^{m+1}}e^{z}|z\rangle,
\end{equation}
where $z$ is integrated around a unit circle in the complex plain. To express $|{\bf m}\rangle$ in this manner,
all we need to do is apply the same identity to each state $|m_j\rangle$ with a corresponding integration variable $z_j$.
To see that Eq. (\ref{zIntegral}) is true, we write the state $|z\rangle$ explicitly and set $z=e^{i\theta}$, which leads to
\begin{eqnarray}\label{zIntegralProof}
|m\rangle&=&\int_0^{2\pi} d\theta\frac{m!}{2\pi}e^{-im\theta}e^{e^{i\theta}}\left[\sum_{m'}\frac{e^{im'\theta}}{m'!}e^{-e^{i\theta}}|m'\rangle\right]\nonumber\\
&=&\sum_{m'}\frac{m!}{m'!}\left[\frac{1}{2\pi}\int_0^{2\pi}d\theta e^{i(m'-m)\theta}\right]|m'\rangle=\sum_{m'}\frac{m!}{m'!}\delta_{m'm}|m'\rangle=|m\rangle.
\end{eqnarray}
For convenience, we write
\begin{equation}
z_k(t)=z_k^{(1)}(t)+\sum_{j=1}^Mz_{kj}^{(0)}(t)z_j,
\end{equation}
where 
\begin{equation}
z_k^{(1)}(t)=\int_0^tf_k(t-t')r(t')dt',\,\,\,\,\,\text{and}\,\,\,\,\,z_{kj}^{(0)}(t)=\sum_{j=i}^MU_{ki}U_{ij}^{-1}e^{S_it}.
\end{equation}
Then,
\begin{eqnarray}\label{Gm}
G(t)|{\bf m}\rangle&=&\prod_{j=1}^M\left[\oint dz_j\frac{m_j!}{2\pi iz_j^{m_j+1}}e^{z_j}\right]G(t)|{\bf z}\rangle\nonumber\\
&=&\prod_{j=1}^M\left[\oint dz_j\frac{m_j!}{2\pi iz_j^{m_j+1}}e^{z_j}\right]|{\bf z}(t)\rangle\nonumber\\
&=&\sum_{\bf m'}\prod_{j=1}^M\left[\oint dz_j\frac{m_j!}{2\pi iz_j^{m_j+1}}e^{z_j}\right]
\frac{\left(z_j^{(1)}(t)+\sum_{l=1}^Mz_{jl}^{(0)}(t)z_l\right)^{m'_j}}{m'_j!}\nonumber\\
&\times&\text{exp}\left[-z_j^{(1)}(t)-\sum_{l=1}^Mz_{jl}^{(0)}(t)z_l\right]|{\bf m}'\rangle.
\end{eqnarray}
The GF for the variable $m_k$ then reads
\begin{eqnarray}
F_k({\bf m},x,t)&=&\prod_{j=1}^M\left[\int_0^{2\pi} d\theta_j\frac{m_j!}{2\pi}e^{-im_j\theta_j}e^{e^{i\theta_j}}\right]
\sum_{{\bf m}'}x^{m_k}\langle{\bf m}'|{\bf z}(t)\rangle\nonumber\\
&=&\prod_{j=1}^M\left[\int_0^{2\pi} d\theta_j\frac{m_j!}{2\pi}e^{-im_j\theta_j}e^{e^{i\theta_j}}\right]
\text{exp}\left[(x-1)\left(z_k^{(1)}(t)+\sum_{l=1}^Mz_{kl}^{(0)}(t)e^{i\theta_l}\right)\right]\nonumber\\
&=&\prod_{j=1}^M\left[\int_0^{2\pi} d\theta_j\frac{m_j!}{2\pi}e^{-im_j\theta_j}e^{\left[(x-1)z_{kj}^{(0)}(t)+1\right]e^{i\theta_j}}\right]
\text{exp}\left[(x-1)z_k^{(1)}(t)\right]
\end{eqnarray}
Invoking the integral identity
\begin{equation}
\int_0^{2\pi} d\theta\frac{m!}{2\pi}e^{-im\theta}e^{ae^{i\theta}}=a^m,
\end{equation}
we obtain
\begin{equation}
F_k({\bf m},x,t)=\prod_{j=1}^M\left[(x-1)z_{kj}^{(0)}(t)+1\right]^{m_j}
\text{exp}\left[(x-1)z_k^{(1)}(t)\right],
\end{equation}
where $m_1,..,m_M$ are the initial protein numbers. For an arbitrary initial probability distribution,
$P({\bf m},0)$, the GF becomes
\begin{equation}
F_k(x,t)=\left[\sum_{\bf m}P({\bf m},0)\prod_{j=1}^M\left[(x-1)z_{kj}^{(0)}(t)+1\right]^{m_j}\right]
\text{exp}\left[(x-1)z_k^{(1)}(t)\right],
\end{equation}
or, in terms of the initial GF, $F(x_1,...,x_M,0)$,
\begin{equation}
F_k(x,t)=F(\phi_1(x,t),...,\phi_M(x,t),0)\text{exp}\left[(x-1)z_k^{(1)}(t)\right],
\end{equation}
where $\phi_j(x,t)=(x-1)z_{kj}^{(0)}(t)+1$.

Unfortunately, variables $n_1,...,n_N$ cannot be decoupled from ${\bf m}$ via the same trick.
However, they can be decoupled in a different way. 
Imagine we are able to observe the evolution, or path, of the variable
$m_M$ in real time. Then the Master equation for ${\bf n}$ is simply  
\begin{eqnarray}\label{CME_protein}
\frac{\partial}{\partial t}P({\bf n},t)&=&Km_M(t)[P(n_1-1)-P]+q[(n_N+1)P(n_N+1)-n_NP]\nonumber\\
&&+\sum_{i=1}^{N-1}b_i[(n_i+1)P(n_i+1,n_{i+1}-1)-n_iP]\nonumber\\
&&+\sum_{i=1}^{N-1}{\bar b}_{i+1}[(n_{i+1}+1)P(n_i-1,n_{i+1}+1)-n_{i+1}P].
\end{eqnarray}
This equation is identical in structure to Eq. (\ref{CME_mRNA}). Hence, starting with some initial set $(n_1,...,n_N)$,
the GF for the variable $n_k$ is
\begin{equation}\label{GF_protein}
{\tilde F}_k({\bf n},y,t)=\prod_{j=1}^N\left[(y-1)w_{kj}^{(0)}(t)+1\right]^{n_j}\text{exp}\left[(y-1)w_k^{(1)}(t)\right].
\end{equation}
We put a tilde over $F$ to distinguish it from the actual GF. 
The functions $w_{kj}^{(0)}(t)$ and $w_k^{(1)}(t)$
are defined as
\begin{equation}
w_k^{(1)}(t)=\int_0^tg_k(t-t')m_M(t')dt',\,\,\,\,\,\,\,\,\,w_{kj}^{(0)}(t)=\sum_{i=1}^NV_{ki}V_{ij}^{-1}e^{T_it},
\end{equation}
where
\begin{equation}
g_k(t-t')=\sum_{i=1}^MV_{ki}V_{i1}^{-1}e^{T_i(t-t')},
\end{equation}
$V_{ij}$ is the unitary matrix that diagonalizes the matrix
\begin{eqnarray}
{\bf T}=
\left[\begin{array}{cccccccccc}
-b_1 & {\bar b}_2 &  &  &  &  & &  \\
b_1 & -(b_2+{\bar b}_2) & {\bar b_3} &  &  &  &  & & &\\
&   & {\huge\text{.}} &   &   &   &   &   &   &\\
&   &                         &   &    & {\huge\text{.}} &   &   &   &\\
&   &                         &   &    &                         &    &  {\huge\text{.}}  &   &\\
&   &   &   &   &   &   &   &   &\\
&   &   &   &   &   &   &   &   &\\
&   &   &   &   &   &   & b_{N-2}  & -(b_{N-1}+{\bar b}_{N-1}) &  {\bar b}_N\\
&   &   &   &   &   &   &  & b_{N-1} &  -(b_N+q)\nonumber
\end{array}\right],
\end{eqnarray}
and $T_i$ is the $i^{th}$ eigenvalue of ${\bf T}$.
The GF (\ref{GF_protein}) is valid only for a particular path taken by the variable $m_M$.
To obtain the true GF,  we must multiply Eq. (\ref{GF_protein}) by the probability of observing
a particular path and then sum over all possible paths:
\begin{eqnarray}\label{GFpath}
F_k({\bf n},y,t)&=&\sum_{\text{all paths}}P({\bf m}(t_0),0){\mathcal P}(\{{\bf m}\}){\tilde F}_k({\bf n},y,t)\nonumber\\
&=&\prod_{j=1}^N\left[(y-1)w_{kj}^{(0)}(t)+1\right]^{n_j}Q(t,t),
\end{eqnarray}
where
\begin{equation}\label{Qdef}
Q(t,t)=\sum_{{\bf m}_0,...,{\bf m}_L}P({\bf m}(t_0),0){\mathcal P}(\{{\bf m}\})
\text{exp}\left[(y-1)\int_0^tg_k(t-t')m_M(t')dt'\right],
\end{equation}
${\mathcal P}(\{{\bf m}\})$ is the probability to observe
a particular set 
\newline
$\{{\bf m}\}=({\bf m}(t_0),{\bf m}(t_1),...,{\bf m}(t_L))$, such that ${\bf m}(t_0)$ is the set of mRNA copy numbers $(m_1,m_2,...,m_M)$ at $t_0=0$,
${\bf m}(t_1)$ is the set of mRNA copy numbers $(m_1,m_2,...,m_M)$ at $t_1=dt$, and so on until $t_L=Ldt=t$; and
$P({\bf m}(t_0),0)$ is the probability of observing the set ${\bf m}(t_0)$ at $t=0$.
We can work out Eq. (\ref{Qdef}) using the following theorem.
\vspace{2mm}
\newline
{\it Theorem 1}: If $P({\bf m},0)$ is the probability to observe ${\bf m}$ at $t=0$
in a system governed by
Eq. (\ref{CME_mRNA}), then, for an arbitrary functional $G({\bf m}(t'),t,t')$,
\begin{equation}\label{exp}
\sum_{\text{all paths}}P({\bf m}(t_0),0){\mathcal P}(\{{\bf m}\})\textrm{exp}\left[\int_0^tG({\bf m}(t'),t,t')dt'\right]
=\sum_{\bf m}Q({\bf m},t,t')\bigg|_{t'=t}
\end{equation}
where $Q({\bf m},t,t')$ is the solution of
\begin{eqnarray}\label{Q_CME_mRNA}
\frac{dQ({\bf m},t,t')}{dt'}&=&
r(t')[Q(m_1-1)-Q]+d[(m_M+1)Q(m_M+1)-m_MQ]\nonumber\\
&&+\sum_{i=1}^{M-1}b_i[(m_i+1)Q(m_i+1,m_{i+1}-1)-m_iQ]\nonumber\\
&&+\sum_{i=1}^{M-1}{\bar b}_{i+1}[(m_{i+1}+1)Q(m_i-1,m_{i+1}+1)-m_{i+1}Q]\nonumber\\
&&+G({\bf m},t,t')Q
\end{eqnarray}
such that $Q({\bf m},t,0)=P({\bf m},0)$.
\vspace{5mm}
\newline
{\it Proof}:
The probability to observe
a path $\{{\bf m}\}$ is given by
\begin{eqnarray}
&&\langle {\bf m}(t_L)|e^{H(t_{L-1})dt}|{\bf m}(t_{L-1})\rangle\langle {\bf m}(t_{L-1})|e^{H(t_{L-2})dt}|{\bf m}(t_{L-2})\rangle
\,.\,.\,.\nonumber\\
&&\,.\,.\,.\,\langle {\bf m}(t_1)|e^{H(t_0)dt}|m(t_0)\rangle.
\end{eqnarray}
Then,
\begin{eqnarray}
&&\sum_{\text{all paths}}P({\bf m}(t_0),0){\mathcal P}(\{{\bf m}\})\textrm{exp}\left[\int_0^tG({\bf m}(t'),t,t')dt'\right]\nonumber\\
&&=\sum_{\text{all paths}}P({\bf m}(t_0),0)\prod_{i=0}^{L-1} e^{G({\bf m}(t_i),t,t_i)dt}\nonumber\\
&&\times\langle {\bf m}(t_L)|e^{H(t_{L-1})dt}|{\bf m}(t_{L-1})\rangle\langle {\bf m}(t_{L-1})|e^{H(t_{L-2})dt}|{\bf m}(t_{L-2})\rangle
\,.\,.\,.\nonumber\\
&&\,.\,.\,.\,\langle {\bf m}(t_1)|e^{H(t_0)dt}|m(t_0)\rangle.
\end{eqnarray}
We can rearrange these products so as to combine the indices in $G({\bf m}(t_i),t,t_i)$ with those labeling the basis, like so
\begin{eqnarray}\label{pathh}
&&\sum_{\text{all paths}}\langle {\bf m}(t_L)|e^{H(t_{L-1})dt}\left[e^{G({\bf m}(t_{L-1}),t,t_{L-1})dt}|{\bf m}(t_{L-1})\rangle\right]\langle {\bf m}(t_{L-1})|\nonumber\\
&&\times e^{H(t_{L-2})dt}\left[e^{G({\bf m}(t_{L-2}),t,t_{L-2})dt}|{\bf m}(t_{L-2})\rangle\right]\langle {\bf m}(t_{L-2})|\,.\,.\,.\,\nonumber\\
&&\,.\,.\,.\,|{\bf m}(t_1)\rangle\langle {\bf m}(t_1)|e^{H(t_0)dt}\left[e^{G({\bf m}(t_0),t,t_0)dt}|{\bf m}(t_0)\rangle\right]P({\bf m}(t_0),0).
\end{eqnarray}
Since $A_i^+A_i|{\bf m}(t_k)\rangle=m_i(t_k)|{\bf m}(t_k)\rangle$, for $i=1,2,...,M$,
we can replace the set ${\bf m}=(m_1,m_2,...,m_M)$ in $G({\bf m}(t_i),t,t_i)$ with the set 
\newline
${\bf A^+A}=(A_1^+A_1,A_2^+A_2,...,A_M^+A_M)$. 
This allows us to move the sums over individual times from the very front of Eq. (\ref{pathh}) to the immediate left of each
basis:
\begin{eqnarray}\label{pathh2}
&&\sum_{{\bf m}(t_L)}\langle {\bf m}(t_L)|e^{H(t_{L-1})dt}e^{G({\bf A^+A},t,t_{L-1})dt}\left[\sum_{{\bf m}(t_{L-1})}|{\bf m}(t_{L-1})\rangle\langle {\bf m}(t_{L-1})|\right]\nonumber\\
&&\times e^{H(t_{L-2})dt}e^{G({\bf A^+A},t,t_{L-2})dt}\left[\sum_{{\bf m}(t_{L-2})}|{\bf m}(t_{L-2})\rangle\langle {\bf m}(t_{L-2})|\right]\,.\,.\,.\,\nonumber\\
&&\,.\,.\,.\,e^{H(t_0)dt}e^{G({\bf A^+A},t,t_0)dt}\sum_{{\bf m}(t_0)}|{\bf m}(t_0)\rangle P({\bf m}(t_0),0),
\end{eqnarray}
where we have rearranged the square brackets to call attention to the identity
\begin{equation}\label{Identity}
\mathbbm{1}=\sum_{\bf m}|{\bf m}\rangle\langle {\bf m}|.
\end{equation}
Hence, Eq. (\ref{pathh2}) becomes 
\begin{eqnarray}\label{pathhh}
&&\sum_{{\bf m}(t_L)}\langle {\bf m}(t_L)|\bigg[e^{H(t_{L-1})dt}e^{G({\bf A^+A},t,t_{L-1})dt}
\,.\,.\,.\,e^{H(t_0)dt}e^{G({\bf A^+A},t,t_0)dt}\bigg]\nonumber\\
&&\times\sum_{{\bf m}(t_0)}|{\bf m}(t_0)\rangle P(({\bf m}(t_0),0).\nonumber\\
\end{eqnarray}
The entire operator in the square brackets now acts on the initial state $|\psi(0)\rangle$ defined in Eq. (\ref{InitialState}).
Invoking the relation $e^{sV}e^{sW}=e^{s(V+W)}+\mathcal{O}(s^2)$ for arbitrary matrices $V$ and $W$ in the limit $s\rightarrow 0$, 
expression (\ref{pathhh}) can be
written as
\begin{eqnarray}\label{pathfinal}
\sum_{{\bf m'}}\left[\langle {\bf m'}|e^{{\tilde H}(t,t_{L-1})dt}e^{{\tilde H}(t,t_{L-2})dt}\,.\,.\,.\,e^{{\tilde H}(t,t_0)dt}|\psi(0)\rangle\right],
\end{eqnarray}
where ${\bf m}$ and ${\bf m'}$ refer to the set $(m_1,m_2,...,m_M)$ at $t'=0$ and $t'=t$, respectively, and ${\tilde H}(t,t')=H(t')+G({\bf A^+A},t,t')$. 
The expression in Eq. (\ref{pathfinal}), satisfies the equation
\begin{equation}
\frac{d}{dt'}|\varphi\rangle={\tilde H}(t,t')|\varphi\rangle,
\end{equation}
or, if we define the state $|\varphi\rangle$ as
\begin{equation}
|\varphi\rangle=\sum_{\bf m}Q({\bf m},t,t')|{\bf m}\rangle,
\end{equation}
$Q({\bf m},t,t')$ must satisfy Eq. (\ref{Q_CME_mRNA}).
For $t'=t_0=0$, the only term in the brackets of Eq. (\ref{pathfinal}) is
$e^{{\tilde H}(t,0)dt}=1+{\tilde H}(t,0)dt+\mathcal{O}(dt^2)$; hence, as $dt\rightarrow 0$, Eq. (\ref{pathfinal}) reduces to
\begin{equation}
\sum_{{\bf m'}}\langle {\bf m'}|\varphi\rangle=\sum_{{\bf m'}}\langle {\bf m'}|\left[1+{\tilde H}(t,0)dt+\mathcal{O}(dt^2)\right |\psi(0)\rangle
\end{equation}
and $|\varphi\rangle\rightarrow|\psi(0)\rangle$, or $Q({\bf m},t,0)=P({\bf m},0)$. QED
\vspace{5mm}
\newline

To evaluate Eq. (\ref{Qdef}), 
we need only to replace $G({\bf m},t,t')$ in Eq. (\ref{Q_CME_mRNA}) with $(y-1)g_k(t-t')m_M$ and solve for $Q({\bf m},t,t')$. 
If the initial probability distribution for mRNA is Poisson,
\begin{equation}\label{psisol2}
P({\bf m},0)=\prod_{i=1}^M\frac{\eta_i^{m_i}}{m_i!}e^{-\eta_i},
\end{equation}
then $Q({\bf m},t,t')$ has the form
\begin{equation}\label{QQ}
Q({\bf m},t,t')=\prod_{i=1}^M\frac{[\xi_i(t,t')]^{m_i}}{m_i!}e^{-h(t,t')}.
\end{equation}
Collecting the factors of $n^1$ and $n^0$, we obtain the equations for ${\pmb\xi}$ and $h$:
\begin{eqnarray}
&&\frac{d{\pmb\xi}}{dt'}={\bf r}+{\bf S}{\pmb\xi}+(y-1)g_k(t-t'){\bf B}{\pmb\xi}\label{xi}\\
&&\frac{dh}{dt'}=r-b\xi_M\label{h},
\end{eqnarray}
where ${\bf B}=\delta_{i,M}\delta_{j,M}$. The dependence of ${\pmb\xi}$ and $h$ on the index $k$ was
left out for the sake of simplicity. To satisfy the initial conditions $Q({\bf m},t,0)=P({\bf m},0)$,
we must have $\xi_i(t,t'=0)=\eta_i$ and $h(t,t'=0)=\sum_i\eta_i$.
Summing Eq. (\ref{QQ}) over ${\bf m}$ leads to
\begin{equation}\label{Q1}
Q(t,t)=\textrm{exp}\left[\sum_{i=1}^M\xi_i(t,t)-h(t,t)\right].
\end{equation}
Adding up all equations in (\ref{xi}) and subtracting Eq. (\ref{h}), we obtain
\begin{equation}\label{sumxih}
\frac{d}{dt'}\left[\sum_{i=1}^M\xi_i(t,t')-h(t,t')\right]=(y-1)g_k(t-t')\xi_M(t,t').
\end{equation}
Integrating Eq. (\ref{sumxih}) over $dt'$, we obtain
\begin{equation}\label{Q2}
Q(t,t)=\textrm{exp}\left[(y-1)\int_0^tg_k(t-t')\xi_M(t,t')dt'\right].
\end{equation}
The former solution to Eq. (\ref{xi}) is
\begin{equation}\label{xiformal}
{\pmb\xi}(t,t')={\bf \tilde{D}}(t,t')+{\bf D}(t,t',0){\pmb\xi}(0),
\end{equation}
where
\begin{eqnarray}
{\bf \tilde{D}}(t,t')&=&\int_0^{t'}{\bf D}(t,t',t''){\bf r}(t'')dt'',\label{D1}\\
{\bf D}(t,t',t'')&=&T\textrm{exp}\left[\int_{t''}^{t'}\left[{\bf S}+(y-1)g_k(t-t_1){\bf B}\right]dt_1\right]\label{D2}
\end{eqnarray}
and $T$ is again the time ordering operator. Inserting ${\bf \tilde{D}}(t,t')$ and ${\bf D}(t,t',0){\pmb\xi}(0)$
into Eq. (\ref{xi}) separately, we obtain one set of equations for the vector  ${\bf \tilde{D}}(t,t')$ and another
set of equations for the matrix ${\bf D}(t,t',0)$:
\begin{eqnarray}
&&\frac{d}{dt'}{\bf \tilde{D}}(t,t')={\bf r}+{\bf S}{\bf \tilde{D}}(t,t')+(y-1)g_k(t-t'){\bf B}{\bf \tilde{D}}(t,t')\label{Dvector}\\
&&\frac{d}{dt'}{\bf D}(t,t',0)={\bf S}{\bf D}(t,t',0)+(y-1)g_k(t-t'){\bf B}{\bf D}(t,t',0)\label{Dmatrix},
\end{eqnarray}
with initial conditions that follow from Eqs. (\ref{D1}) and (\ref{D2}): ${\bf \tilde{D}}(t,0)=0$ and ${\bf D}(t,0,0)=\mathbbm{1}$.
Eq. (\ref{Q2}) can now be written as
\begin{equation}\label{Q3}
Q(t,t)=\textrm{exp}\left\{(y-1)\left[\psi_k^{(1)}(t)+\sum_{j=1}^M\psi_{kj}^{(0)}(t)\xi_j(0)\right]\right\},
\end{equation}
where
\begin{equation}\label{psik1}
\psi_k^{(1)}(t)=\int_0^tg_k(t-t'){\tilde{D}}_M(t,t')dt'
\end{equation}
and
\begin{equation}\label{psik2}
\psi_{kj}^{(0)}(t)=\int_0^tg_k(t-t')D_{Mj}(t,t',0)dt'.
\end{equation}
Note that in Eqs. (\ref{Q3}), (\ref{psik1}) and (\ref{psik2}) we put back the index $k$.

Results in Eqs. (\ref{Q3}), (\ref{psik1}) and (\ref{psik2}) are correct only if the initial probability distribution, $P({\bf m},0)$, is Poisson.
To obtain $Q(t,t)$ for an arbitrary $P({\bf m},0)$, we can apply the same trick as in
Eq. (\ref{Gm}), except that now the operator $G(t)$ must be replaced with
\begin{equation}\label{psisol2}
{\tilde G}(t)=T\text{exp}\left[\int_0^t[H(t')+(y-1)g_k(t-t')m_M]dt'\right].
\end{equation}
The action of ${\tilde G}(t)$ on a Poisson state $|{\bf z}\rangle$, where ${\bf z}$ is any complex number,
produces a new state
\begin{equation}
{\tilde G}(t)|{\bf z}\rangle=|{\bf z}(t)\rangle=\sum_{\bf m}\prod_{i=1}^M\frac{[\xi_i(t,t)]^{m_i}}{m_i!}e^{-h(t,t)}|{\bf m}\rangle,
\end{equation}
with the initial conditions $\xi_i(t,0)=z_i$ and $h(t,0)=\sum_iz_i$.
Invoking Eq. (\ref{zIntegral}), we can write the expression for $Q(t,t)$, for the initial conditions
$Q({\bf m}',t,0)=\delta_{m'_1,m_1}...\delta_{m'_M,m_M}$, as
\begin{eqnarray}\label{Qfinal}
Q(t,t)&=&\sum_{{\bf m}'}\langle {\bf m'}|{\tilde G}(t)|{\bf m}\rangle\nonumber\\
&=&\prod_{j=1}^M\left[\int_0^{2\pi} d\theta_j\frac{m_j!}{2\pi}e^{-im_j\theta_j}e^{e^{i\theta_j}}\right]
\sum_{{\bf m}'}\langle{\bf m}'|{\bf z}(t)\rangle\nonumber\\
&=&\prod_{j=1}^M\left[\int_0^{2\pi} d\theta_j\frac{m_j!}{2\pi}e^{-im_j\theta_j}e^{e^{i\theta_j}}\right]
\text{exp}\left[(y-1)\left(\psi_k^{(1)}(t)+\sum_{l=1}^M\psi_{kl}^{(0)}(t)e^{i\theta_l}\right)\right]\nonumber\\
&=&\prod_{j=1}^M\left[\int_0^{2\pi} d\theta_j\frac{m_j!}{2\pi}e^{-im_j\theta_j}e^{\left[(y-1)\psi_{kj}^{(0)}(t)+1\right]e^{i\theta_j}}\right]
\text{exp}\left[(y-1)\psi_k^{(1)}(t)\right]\nonumber\\
&=&\prod_{j=1}^M\left[(y-1)\psi_{kj}^{(0)}(t)+1\right]^{m_j}\text{exp}\left[(y-1)\psi_k^{(1)}(t)\right].
\end{eqnarray}
Finally, inserting $Q(t,t)$ in Eq. (\ref{Qfinal}) into Eq. (\ref{GFpath}), we obtain
\begin{eqnarray}
F_k({\bf m},{\bf n},y,t)&=&\left[\prod_{j=1}^N\left[(y-1)\psi_{kj}^{(0)}(t)+1\right]^{m_j}\left[(y-1)w_{kj}^{(0)}(t)+1\right]^{n_j}\right]\nonumber\\
&\times&\text{exp}\left[(y-1)\psi_k^{(1)}(t)\right].
\end{eqnarray}
For an arbitrary initial joint probability distribution, $P({\bf m},{\bf n},0)$, the GF reads
\begin{eqnarray}\label{FinalResult}
F_k(y,t)&=&\left[\sum_{\bf m}\sum_{\bf n}P({\bf m},{\bf n},0)\prod_{j=1}^N\left[(y-1)\psi_{kj}^{(0)}(t)+1\right]^{m_j}
\left[(y-1)w_{kj}^{(0)}(t)+1\right]^{n_j}\right]\nonumber\\
&\times&\text{exp}\left[(y-1)\psi_k^{(1)}(t)\right]\nonumber\\
&=&F(\mu_1(t),...,\mu_M(t),\nu_1(t),...,\nu_N(t),0)\text{exp}\left[(y-1)\psi_k^{(1)}(t)\right],
\end{eqnarray}
where $F(u_1,...,u_{M+N},0)$ is the initial GF, $\mu_i(t)=(y-1)\psi_{kj}^{(0)}(t)+1$ and
$\nu_i(t)=(y-1)w_{kj}^{(0)}(t)+1$.

\section{Results}

In this section we compute probability distributions and central moments using result (\ref{FinalResult})
and compare them to Gillespie simulations. 

The probability of observing $n_N$ fully processed proteins is given in Eq. (\ref{PF}). 
To take the $n^{\text{th}}$ derivative of $F$, we apply Cauchy's integral formula, which states that
\begin{equation}\label{Cauchy}
\frac{d^nf(y)}{dx^n}=\frac{n!}{2\pi i}\oint dz\frac{f(z)}{(z-y)^{n+1}},
\end{equation}
where $f(x)$ is a function that is analytic at the point $y$. 
The integral over the complex variable $z$ must enclose the point $y$ but is
otherwise arbitrary. We chose the contour to be
a unit circle in the complex plain, i. e. $z=e^{i\theta}$,
solve numerically Eqs. (\ref{Dvector}) and (\ref{Dmatrix})
for $y=e^{i2\pi j/J}$ for every $j=0,1,...,J$, with $J=100$, and interpolate the solutions with a quadratic spline.
Then, we perform the Cauchy integral for $y=0$.

To obtain the first four statistical moments, we take successive derivatives of Eqs. (\ref{Dvector}) and (\ref{Dmatrix})
with respect to $y$, and set $y=1$. This leads to the following sets of equations: 
\begin{eqnarray}
&&\frac{d}{dt'}{\bf \tilde{D}}^{(0)}(t,t')={\bf r}+{\bf S}{\bf \tilde{D}}^{(0)}(t,t')\nonumber\\
&&\frac{d}{dt'}{\bf \tilde{D}}^{(i)}(t,t')={\bf S}{\bf \tilde{D}}^{(i)}(t,t')+g_k(t-t'){\bf B}{\bf \tilde{D}}^{(i-1)}(t,t')\,\,\,\,\,\,\,\,\,\,\,\,\,\,\,\,\,\,\,\,\,\,\,0<i\nonumber\\
&&\frac{d}{dt'}{\bf D}^{(0)}(t,t',0)={\bf S}{\bf D}^{(0)}(t,t',0)\nonumber\\
&&\frac{d}{dt'}{\bf D}^{(i)}(t,t',0)={\bf S}{\bf D}^{(i)}(t,t',0)+g_k(t-t'){\bf B}{\bf D}^{(i-1)}(t,t',0)\,\,\,\,\,\,0<i,\nonumber\\
\end{eqnarray}
where the upper index stands for the order of the derivative.
Then, considering only the first four moments, $\psi_k^{(1)}(t)$ and $\psi_{kj}^{(0)}(t)$ can be written as polynomials
on $(y-1)$:
\begin{eqnarray}
\psi_k^{(1)}(t)&=&\sum_{n=0}^3(y-1)^{n+1}\left[\int_0^tg_k(t-t'){\tilde{D}}_M^{(n)}(t,t')dt'\right]\\
\psi_{kj}^{(0)}(t)&=&\sum_{n=0}^3(y-1)^{n+1}\left[\int_0^tg_k(t-t')D_{Mj}(t,t',0)^{(n)}dt'\right].
\end{eqnarray}
With respect to $y$, the GF in Eq. (\ref{FinalResult}) is now analytic.
Taking the first four derivatives of Eq. (\ref{FinalResult}) and setting $y=1$, we obtain the following 
combinations of statistical moments for the variable $n_N$:
\begin{eqnarray}\label{theF}
F^{(1)}&=&\langle n_N\rangle\nonumber\\
F^{(2)}&=&\langle n_N(n_N-1)\rangle=\langle n_N^2\rangle-\langle n_N\rangle\nonumber\\
F^{(3)}&=&\langle n_N(n_N-1)(n_N-2)\rangle=\langle n_N^3\rangle-3\langle n_N^2\rangle+2\langle n_N\rangle\nonumber\\
F^{(4)}&=&\langle n_N(n_N-1)(n_N-2)(n_N-3)\rangle=\langle n_N^4\rangle-6\langle n_N^3\rangle+11\langle n_N^2\rangle-6\langle n_N\rangle\nonumber\\
\end{eqnarray}
where $F^{(l)}=[d^lF_N(y,t)/(dy)^l]_{y=1}$. The first four central moments, defined as $\sigma_1=\langle n_N\rangle$, $\sigma_l=\langle(n-\langle n\rangle)^l\rangle$ for $l=2,3...$,
are
\begin{eqnarray}\label{CentralMoments}
\sigma_1&=&\langle n_N\rangle\nonumber\\
\sigma_2&=&\langle n_N^2\rangle-\langle n_N\rangle^2\nonumber\\
\sigma_3&=&\langle n_N^3\rangle-3\langle n_N\rangle\langle n_N^2\rangle+2\langle n_N\rangle^3\nonumber\\
\sigma_4&=&\langle n_N^4\rangle-4\langle n_N\rangle\langle n_N^3\rangle+6\langle n_N\rangle^2\langle n_N^2\rangle-3\langle n_N\rangle^4.
\end{eqnarray}
Solving Eqs. (\ref{theF}) for the moments $\langle n_N^l\rangle$ and inserting the solution into Eqs. (\ref{CentralMoments}), we obtain
\begin{figure}
\centering
\includegraphics[trim=0 0 0 1.0cm, height=0.7\textheight]{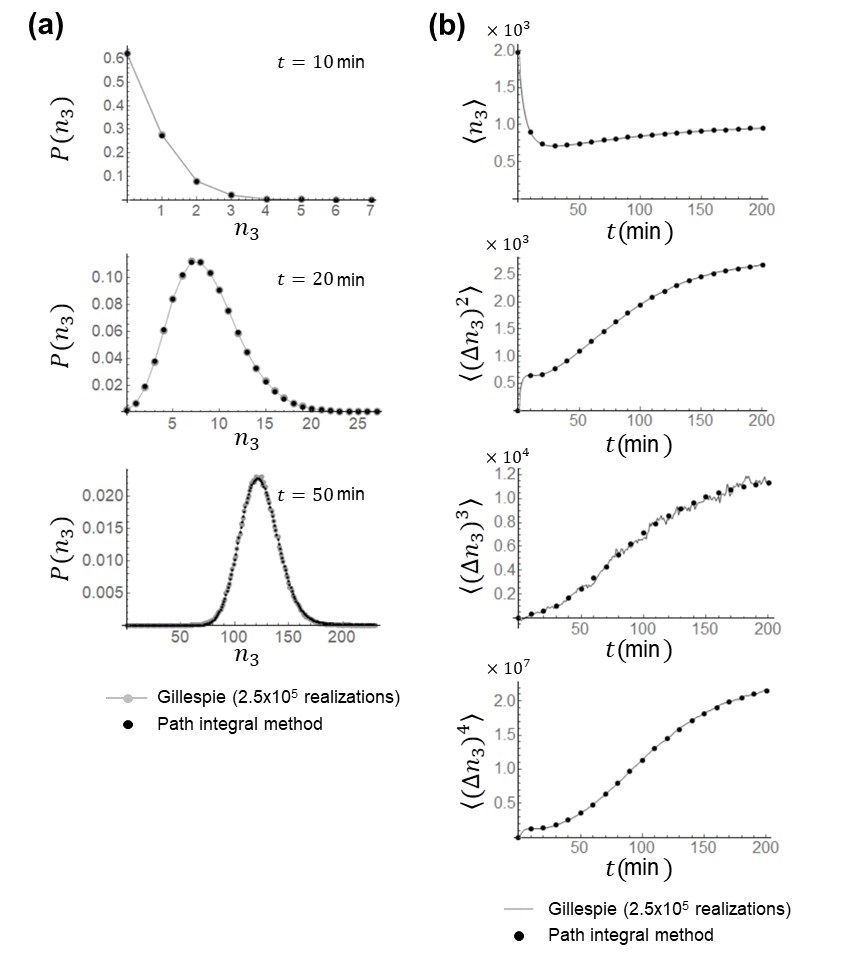}
\caption{Comparison of the path integral method and Gillespie simulations for a system comprising of three 
types of mRNA, three types of protein (M=N=3) and the following reaction rates: $r=5$/min,
$d=0.1$/min, $K=1$/min, $a_i={\bar a}_i=0.2$/min, $b_i={\bar b}_i=0.1$/min and
$q=0.05$/min.
a) probability distribution for $n_3$ at
$t=10, 20$ and $50$min, such that at $t=0$ the system contains zero gene products. b) the first four
central moments, such that at $t=0$, $m_1=59$, $m_2=88$, $m_3=44$, $n_1=982$, $n_2=316$ and $n_3=1977$.}
\end{figure}
\begin{eqnarray}\label{CentralMomentsSol}
\sigma_1&=&F^{(1)}\nonumber\\
\sigma_2&=&F^{(2)}+F^{(1)}-\left[F^{(1)}\right]^2\nonumber\\
\sigma_3&=&F^{(3)}+3F^{(2)}-3F^{(1)}\left[F^{(1)}+F^{(2)}\right]+F^{(1)}+2\left[F^{(1)}\right]^3\nonumber\\
\sigma_4&=&F^{(4)}+6F^{(3)}+7F^{(2)}+F^{(1)}-4F^{(1)}\left[F^{(1)}+3F^{(2)}+F^{(3)}\right]\nonumber\\
&+&6\left[F^{(1)}\right]^2\left[F^{(1)}+F^{(2)}\right]-3\left[F^{(1)}\right]^4.
\end{eqnarray}

We test the calculations above on a system with three post-transcription $(M=3)$ and post-translation processes $(N=3)$, and
with rates for the forward reactions that are equal to the rates for the backward reactions, i. e. $a_i={\bar a}_i$
and $b_i={\bar b}_i$ for $a_i\neq b_i$, and for $r(t)=$ a constant. Fig. 1 shows the comparison between our results and Gillespie simulations.
Fig. 1 a) shows the probability distributions for three different times, starting with the initial conditions
$P({\bf m},{\bf n},0)=\delta_{m_1,0}\delta_{m_2,0}\delta_{m_3,0}\delta_{n_1,0}\delta_{n_2,0}\delta_{n_3,0}$.
Fig. 1 b) shows the first four central moments for the initial conditions
$P({\bf m},{\bf n},0)=\delta_{m_1,m'_1}\delta_{m_2,m'_2}\delta_{m_3,m'_3}\delta_{n_1,n'_1}\delta_{n_2,n'_2}\delta_{n_3,n'_3}$,
where the set $(m'_1,m'_2,m'_3,n'_1,n'_2,n'_3)$ was chosen randomly (see captions in Fig. 1).

\section{Conclusion}

We presented a novel approach to solving the Master equation for a one gene system
comprising of partially and fully processed mRNA and protein. The success of this approach
lies in the fact that the mRNAs influence the production of the proteins but not vice versa, and in a 
theorem which allows for the conversion of an integral over infinitely many paths taken by the mRNAs 
into an equation similar to the Master equation for the mRNAs. 
A direct approach to solving the Master equation necessitates
the consideration of all variables at once. The effect of our approach is the separation of
the variables, which reduces the dimension of the problem to one. Although we only focused on
probability distributions for each variable separately, the steps detailed in the main text allow for the calculation
of joint probability distributions of two or more variables.
Since the above-mentioned theorem
is true for any system in which one part of the system, A, affects another part, B, unidirectionaly, this approach
can be applied to problems involving several genes. The effect of our approach in such a case would
be the reduction of the problem's dimension from A+B to A.

\section{Acknowledgments}
This work was not funded by any institution or organization.
JA thanks Ekaterina Ejkova for her technical support.



\begin{thebibliography}{10}

\bibitem{Blake} Blake WJ, K\ae{}rn M, Cantor CR, Collins JJ, (2003) 
Noise in eukaryotic gene expression.
Nature volume 422, pages 633–637

\bibitem{Gillespie} Gillespie DT, (1977) 
Exact Stochastic Simulation of Coupled Chemical Reactions. J.
Phys. Chem. 81(25), 2340-2361

\bibitem{Gibson} Gibson MA, Bruck J, (2000) 
Efficient Exact Stochastic Simulation of Chemical Systems with Many Species and Many Channels. 
J. Phys. Chem. 104(9), 1876–1889

\bibitem{Gillespie2} Gillespie DT, (2001)
Approximate accelerated stochastic simulation of chemically reacting systems. 
J. Chem. Phys. 115(4), 1716

\bibitem{Cao} Cao Y, Li H, Petzold L, (2004) 
Efficient formulation of the stochastic simulation algorithm for chemically reacting systems. 
J. Chem. Phys. 121, 4059

\bibitem{Cao2} Cao Y, Gillespie DT, Petzold LR, (2005) 
Avoiding negative populations in explicit Poisson tau-leaping. J. Chem. Phys. 123(5), 054104

\bibitem{Cao3} Cao Y, Gillespie DT, Petzold LR, (2005)
Efficient step size selection for the tau-leaping simulation method. J. Chem. Phys. 124(4), 044109

\bibitem{Burrage} Burrage K, Tian T, Burrage P, (2004)
A multi-scaled approach for simulating chemical reaction systems. 
Progress in Biophysics \& Molecular Biology, 85, 217-234


\bibitem{Jahnke} Jahnke T, Altıntan D, (2010)
Efficient simulation of discrete stochastic reaction systems with a splitting method. 
BIT Num Math 50(4), 797-822

\bibitem{Albert} Albert J, (2016)
A hybrid of the chemical master equation and the Gillespie algorithm for efficient stochastic simulations of sub-networks.
PloS one 11 (3), e0149909

\bibitem{Albert12} Albert J, (2016)
Stochastic simulation of reaction subnetworks: Exploiting synergy between the chemical master equation and the Gillespie algorithm
AIP Conference Proceedings 1790 (1), 150026

\bibitem{Duso} Duso L, Zechner C, (2018)
Selected-node stochastic simulation algorithm
J. Chem. Phys, 148, 164108 

\bibitem{Alfonsi} Alfonsi A, Cances E, Turinic G, Ventura BD, Huisinga W, (2005)
Adaptive simulation of hybrid stochastic and deterministic models for biochemical systems. 
ESAIM: Proc. 14, 1-13

\bibitem{Kurasov} Kurasov P, L\"{u}ck A, Mugnolo D, Wolf V, (2018)
Stochastic Hybrid Models of Gene Regulatory Networks
arXiv preprint arXiv:1803.10958

\bibitem{Jahnke2} Jahnke T, Huisinga W, (2007)
Solving the chemical master equation for monomolecular reaction systems analytically.
J Math Biol. 54(1):1-26

\bibitem{Albert2} Albert J, Rooman M, (2016)
Probability distributions for multimeric systems
J. math. biol. 72 (1-2), 157-169

\bibitem{Shahrezaei} Shahrezaei V, Swain PS, (2008)
Analytical distributions for stochastic gene expression.
PNAS, 105(45): 17256–17261.

\bibitem{Pendar} Pendar H, Platini T, Kulkarni RV, (2013)
Exact protein distributions for stochastic models of gene expression using partitioning of Poisson processes
Phys. Rev. E, 87, 042720

\bibitem{Walczak} Aleksandra M. Walczak, Andrew Mugler, Chris H. Wiggins, (2012)
Analytic Methods for Modeling Stochastic Regulatory Networks
Computational Modeling of Signaling Networks 880, 273-322



\end{thebibliography}
\end{document}